\documentclass{sig-alternate}
\usepackage{mathptmx} % This is Times font

\usepackage{fancyhdr}
\usepackage[normalem]{ulem}
\usepackage[hyphens]{url}
\usepackage[sort,nocompress]{cite}
\usepackage[final]{microtype}
\usepackage[bookmarks=true,breaklinks=true,letterpaper=true,colorlinks,linkcolor=black,citecolor=blue,urlcolor=black]{hyperref}
\usepackage{subfig}
\usepackage{placeins}
\usepackage{float}

\graphicspath{ {./figures/} }

\begin{document}

\numberofauthors{3} %  in this sample file, there are a *total*
\author{
% 1st. author
\alignauthor
Adam Hastings\\
       \affaddr{Columbia University}\\
       \affaddr{New York, New York}\\
       \email{hastings@cs.columbia.edu}
% 2nd. author
\alignauthor
Lydia B. Chilton\\
       \affaddr{Columbia University}\\
       \affaddr{New York, New York}\\
       \email{chilton@cs.columbia.edu}
% 3rd. author
\alignauthor
Simha Sethumadhavan\\
       \affaddr{Columbia University}\\
       \affaddr{New York, New York}\\
       \email{simha@columbia.edu}
}

\pagenumbering{arabic}

%%%%%%%%%%%---SETME-----%%%%%%%%%%%%%
\title{How Much is Performance Worth to Users?\\A Quantitative Approach} 
%%%%%%%%%%%%%%%%%%%%%%%%%%%%%%%%%%%%

\maketitle
\pagestyle{plain}

\sloppy

%%%%%% -- PAPER CONTENT STARTS-- %%%%%%%%

\begin{abstract}
    Architects and systems designers artfully balance multiple competing design constraints during the design process but are unable to translate between system metrics and end user experience. 
    This work presents three methodologies to fill in this gap. 
    The first is an incentive-compatible methodology that determines a ``ground truth'' measurement of users' value of  speed in terms of US dollars, and find that users would accept a performance losses of 10\%, 20\%, and 30\% to their personal computer in exchange for \$2.27, \$4.07, and \$4.43 per day, respectively.
    However, while highly accurate the methodology is a painstaking process and does not scale with large numbers of participants. 
    To allow for scalability, we introduce a second methodology---a lab-based simulation experiment---which finds that users would accept a permanent performance loss of 10\%, 20\%, and 30\% to their personal computer in exchange for \$127, \$169, and \$823, respectively.
    Finally, to allow for even greater scalability, we introduce a third methodology---a survey---and observe that the lack of incentive compatibility and the lack of hands-on experience with throttled device performance skews the results significantly, thus demonstrating the need for lab-based or incentive compatible study designs. 
    By quantifying the tradeoff between user satisfaction and performance, we enable architects and systems designers to make more nuanced tradeoffs between design requirements.
\end{abstract}

\section{Introduction} \label{sec:intro}

A quantitative approach to computer architecture has been a key
factor in achieving meaningful improvements for several decades
now.  
Computer architects have done a tremendous job at developing metrics
to measure important quantities such as performance, power consumption,
die area, and reliability, which have helped us have a clear
conversation about pros and cons of competing approaches.  
More recently, computer architects are increasingly tasked with designing
systems for ``soft'' user-facing requirements such as responsiveness,
security, and privacy, which can be broadly classified as a quality
requirements ~\cite{intel2020athena, soasta, hoxmeier2000system, galletta2004web}.
Unlike traditional metrics, it has been a challenge to come up with
metrics that architects can use in their studies when
evaluating tradeoffs in the context of these requirements.
Consequently, most of the factors are motivated using bespoke (and
often qualitative) requirements.

In this paper we propose metrics and methods for incorporating soft
requirements in architectural studies in a quantitative manner.
The key idea is to come up with an ``exchange rate'' between user
satisfaction and system performance.  Specifically, we find how
much users would have to be paid in order to be willing to accept
losses to their computer's performance.

By defining this ``exchange rate'' between performance and user satisfaction, we provide systems designers and architects with a quantitative metric by which to balance tradeoffs between performance and other features which may ``cost'' performance, such as security or usability features. 
As an example, consider the following scenario:
Suppose that a certain vulnerability in a product enables ransomware that costs users an average of \$1000 per year, and that the average device lifespan is two years.
Suppose that computer architects design a defense against the vulnerability, but the defense incurs a 30\% overhead.
Is this a worthwhile tradeoff?
Our results suggest no:
According to our results, users require at least \$4.43 per day to accept a 30\% slowdown, or \$3241.20 across a two-year span, which to the average user is a \textit{higher} cost than the expected \$2000 losses due to ransomware over the same two-year span.
Now, suppose that computer architects improve the defense so that it only incurs a 10\% overhead.
According to our results, such a defense ``costs'' users an average of \$2.27 per day, or \$1657.10 over a two year span.
The cost of this overhead is thus \textit{less} than cost of expected ransomware attacks, and hence this new defense provides more protection than it costs.
By eliciting users' value of performance, our work provides a framework for determining a quantitative basis on which to make this and other such decisions.

To elicit users' willingness to accept performance losses, we present
three experimental methodologies with different attractive features
and shortcomings and compare them.  These methodologies are somewhat
analogous to how architectural ideas can be evaluated using different
methodologies such as analytical models, various flavors of simulation,
FPGA prototyping or ASIC prototyping~\cite{das2017prototype}.

The first approach is an incentive compatible methodology, meaning that
participants are incentivized to make decisions and answer questions
according to their true preferences.  This experiment asks participants
to install a computer program on their personal devices that offers
choices between computer performance and money with real-life
consequences: Participants who choose to accept the offered money
over device performance must then actually endure throttled device
performance for several days at a time.  By making our experimental
design incentive compatible, we can establish a ``ground truth''
measurement of users' willingness to accept performance losses.
While very compelling, the incentive compatible experiment
cannot be completed quickly and it can be difficult to enroll users
in the study.

The second method we design forgoes the incentive compatibility
 of the first methodology but still
subjects participants to throttled performance on participants'
personal devices for a  short duration of time (about 15 minutes)
and asks them to perform specific tasks on their computer.  While
it is easier to enroll more people in the study due to the short
duration, there is a certain amount of experimental bias that can
be introduced because of the nature of the tasks that users have
to carry out on their computers.

Less tedious still is our third methodology, which forgoes on-device
demonstrations of throttled device performance altogether in favor
for a survey instrument, and can be quickly and scalable deployed
with large sample sizes.  While studies have shown that the ``wisdom
of crowds" phenomenon to be able to have some predictive power, the
risk is that users who have not experienced actual slowdowns may not
be properly calibrated to provide the correct value of performance
slowdowns.

By conducting all three methodologies, we can then compare results.
The incentive compatible experiment finds that users would trade performance losses of 10\%, 20\%, and 30\% to their personal devices in exchange for a daily payment of \$2.27, \$4.07, and \$4.44 per day, respectively. 
The simulation study finds that users would trade a permanent performance loss of 10\%, 20\% and 30\% to their personal devices in exchange for total payments of \$127, \$169, and \$823, respectively.
By comparison, the survey study finds that users would trade performance losses of 10\%, 20\%, and 30\% in exchange for a daily payment of \$11.32, \$25.26, and \$27.32 and a total payment of $\$499$, $\$1214$, $\$3723$  suggesting that users are not able to properly gauge their value of performance in the absence of incentive compatibility and hands-on experience.

The rest of the paper is organized as follows:
We describe the methodology of our three experiments in Section~\ref{sec:methods} and present results in Section~\ref{sec:results}.  
This is followed by a discussion on some of the results in Section~\ref{sec:discussion}.  
We then outline some of the limitations of our methodology and results in
Section~\ref{sec:limitations}.  
Applications of this work are considered in Section~\ref{sec:applications}, and related work is reviewed in
Section~\ref{sec:rw}.
Finally, this paper concludes in
Section~\ref{sec:conclusion}.
\section{Methods} \label{sec:methods}

% Maybe we should present the incentive compatible version last?

We now present three methodologies to price the value of performance losses. 
All three methodologies are designed to elicit users' willingness to accept (WTA) various degrees of performance losses in terms of dollars.
The WTA---a concept borrowed from economics---is the minimum amount of money that some user would have to be receive in order to accept some fixed-percentage loss in their device's performance.

Participants for all studies were recruited from Mechanical Turk and were required to be at least 18 years old and working in the United States. 
All experimental protocols were reviewed and approved by our IRB. 

\begin{figure*}[t!]
    \centering
    \subfloat[SPECspeed 2017 Integer\label{specspeed}]{%
    \includegraphics[width=0.32\linewidth]{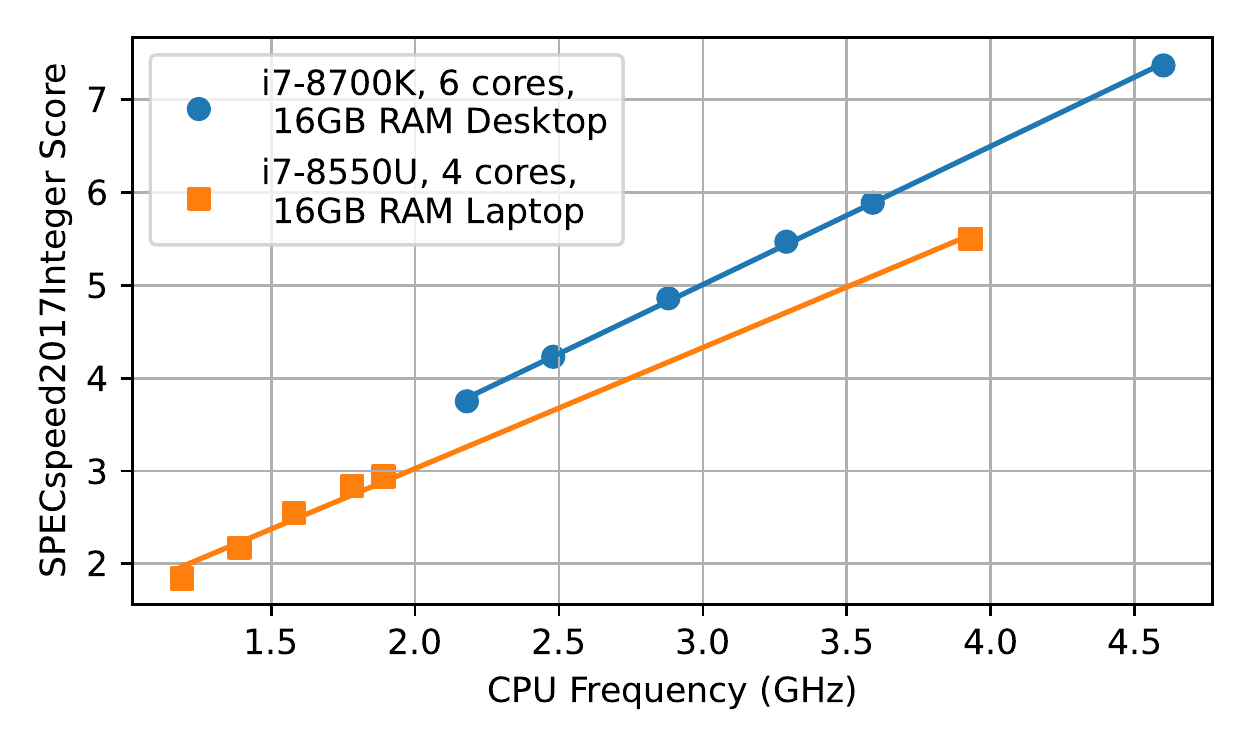}}
    \hfill
    \subfloat[SPECrate 2017 Integer\label{specrate}]{%
    \includegraphics[width=0.32\linewidth]{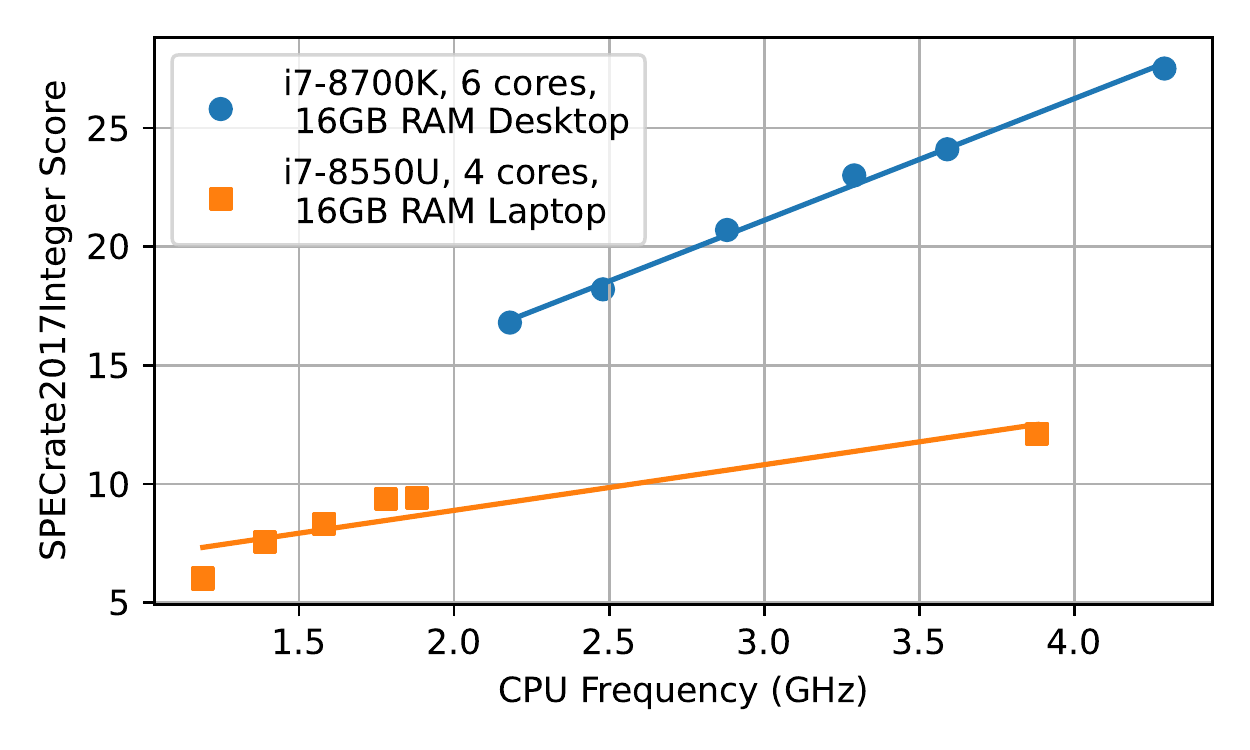}}
    \hfill
    \subfloat[WebXPRT 3\label{webxprt}]{%
    \includegraphics[width=0.32\linewidth]{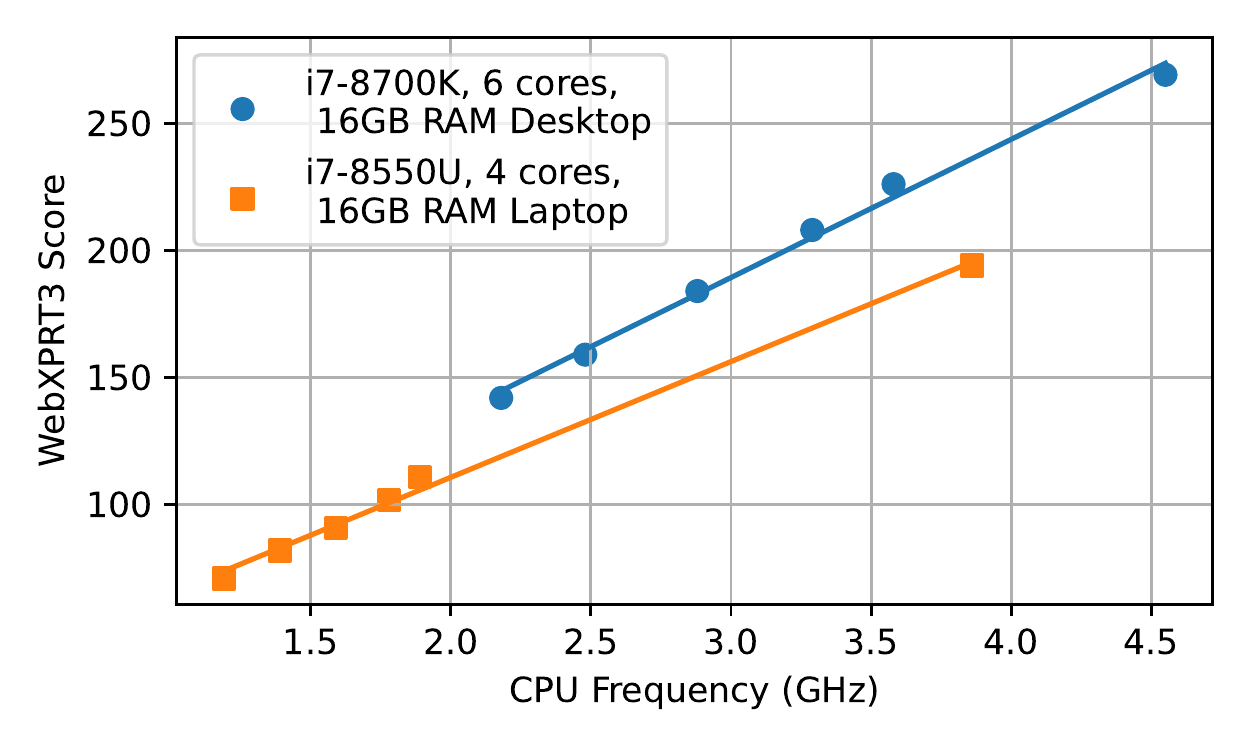}}
    \caption{
        Clock speed strongly correlates with various benchmark scores, suggesting that lowering participants' CPU frequency induces a commensurate loss to their device's performance. We fit a linear trendline to illustrate the high degree correlation between frequency and benchmark scores; Pearson correlation coefficients are above $0.996$ for all datasets except for the SPEC2017 intrate benchmark on the i7-8550U laptop, which had a Pearson correlation coefficient of $0.918$.
    }
    \label{fig:benchmarks} 
\end{figure*}

\subsection{Experiment \#1: Incentive Compatible Study}

% Our first method is a ``ground truth'' measurment 

% something about why incentive compatibility is important
One may ask, why do we need an experimental methodology to determine how much users value performance when we could ask users directly via a survey?
We contend that surveys are insufficient for two reasons:
First, study participants are unlikely to be able to accurately price something they don't understand.
For example, asking study participants to determine how much they would have to be paid to accept a 10\% performance loss to their device is a meaningless question to ask if the participant doesn't understand the implications of what a 10\% performance loss might do to their device.
And second, survey-based methodologies generally lack incentive compatibility, meaning that participants are not incentivized to answer survey questions according to their true preferences (since survey responses have to consequence to the participant) and participants may therefore not put much thought into survey responses, potentially invalidating any results.

To avoid these pitfalls, we design our first experiment to be a  ``ground truth'' measurement:
It is an incentive-compatible methodology where participants are asked to make real-life tradeoffs between money and performance losses on their personal devices. 
This addresses 1) the issue of participants not understanding the implications of performance losses, and 2) the lack of incentive compatibility in survey-based methodologies.

\subsubsection{Experimental Protocol}

The experiment begins with the participant downloading and running a computer program provided by the researchers.
The program first gains consent from the participant, and also asks participants 1) if the program they are using is their primary computing device, and 2) the average number of hours each week that they use the current device.
Participants who either are not using their primary device or do not use their devices for at least 10 hours a week are removed the study to ensure that any potential participants would face actual consequences if the device running the program were to be slowed down.
After this, participants are asked to consent to having their device's clock speed monitored throughout participation (so that the researchers can verify that any induced performance losses have actually occurred) and are asked to consent to using the device at least three-quarters of their average weekly hours as previously reported (also to ensure that participants face actual consequences if the device is slowed down).

Next, participants are given the opportunity to accept some dollar amount $\$P$ (the exact amount varies by participant) in exchange for a performance loss 10\%, 20\%, or 30\% for 24 hours\footnote{We choose to test up to 30\% since this mirrors the worst-case outcomes as seen from some recent hardware security patches. Note that this experiment can be conducted for any slowdown percentage.}.
If the participant declines the offer, they are compensated for their time, and their participation in the study concludes.
However, if the participant chooses to accept the money, the computer program will throttle the CPU's max speed by the agreed upon amount by setting a cap on the participant's device's CPU clock speed (for example, to achieve a 10\% slowdown, we cap participants' CPU frequency at 90\%).
The program then sleeps for 24 hours.

After 24 hours, the program wakes itself up and again offers participants the chance to either restore full performance or accept another $\$P$ in exchange for yet another 24 hours of throttled performance.
This process continues until either the participant eventually declines to accept the money, or until the experiment times out (7 days for the 10\% and 20\% slowdowns and 14 days for the 30\% slowdown).
At this point, performance is fully restored and participants are compensated the small participatory compensation amount plus any additional earnings they have accrued from choosing money over performance.
Once a participant either rejects an offer or reaches times out, they are asked to complete an exit survey.
This concludes the experimental protocol.

The next two subsections describe the steps we take to ensure the validity of the data collected in the above protocol.

\subsubsection{Removing Invalid Results}

% The goal of this experiment is to find the population's threshold dollar amount at which accepting the offer is more likely than not accepting.

After experimentation concludes, participants can be grouped into one of two categories:
Those who accepted the monetary offer for the full duration of the study, and those who did not. 
Those who accept the full offer clearly value the offered money more than the consequences of a throttled device.
In contrast, those who decline the full offer may do so because the offered money was not high enough, but they may also decline for a number of reasons not related to the offer price.
For example, a participant may not trust the researchers to properly slow down their device, or perhaps share their device with others users who may not want to participate in such a longitudinal study.
Since these reasons are extraneous to such participants' willingness to accept performance losses, we ask these ``decliners'' why they chose to decline the given offer in an exit survey.
We consider the data of participants who report to have declined the offer \textit{solely} because the offer price was too low, and discard the data of participants who declined the offer for any other reason.

Another threat to data validity is that some clever participants may try to ``cheat'' the experiment.
For example, a cheating participant may accept the slowdown for the money but then later restore their device's performance manually.
To detect any such cheaters, our program periodically audits device CPU frequency to ensure that CPU frequency has not been tampered with.
Even if such cheating goes undetected, we also ask participants if they cheated during the exit survey and offer amnesty (meaning we promise to issue payments even if they did).
% Second, at the end of participation, participants are asked to report if they cheated or tried to circumvent the slowdown imposed on their device.
% To encourage honesty, we offer immunity (meaning we still promise to issue payments) to participants who admit to cheating.
No cheating was self-reported, but we did detect a few cases of cheating by looking at the audits, and removed these participants' data from further consideration.

Participants may also try to cheat by installing the program on an un- or under-utilized device in an attempt to earn payments without truly enduring a slowdown on their primary desktop device. 
We mitigate this first by informing participants during the consent process that the program will sporadically measure computer clock frequency (and will detect such behavior), and second, by having the program collect timestamps upon waking and analyzing the timestamps for anomalies\footnote{For example, if the participant reports that they use device for 60 hours a week but the timestamps are collected only every few days, then clearly the participant is not actually using their device for 60 hours a week, and would be removed from the study.}.
Thus any participants who are clearly un- or under-utilizing their device are removed from consideration.

Finally, for our results to be valid, participants who accept slowed-down devices in exchange for money must actually endure lower performance.
We ensure this by having the program periodically sample device frequency and report these logs back to our server. 
Because of the difficulty of precisely tuning the frequency of participants' computers, some participants endured slowdowns of greater than the agreed-upon amount of 10\%, 20\%, or 30\% during their participation. 
Since the degree of performance throttling may affect participants' decision making, we removed from consideration any participants who were subjected to substantially higher performance losses than the agreed-upon amount and also initially accepted their given offer only to later reject on a later date.
Conversely, we do not remove from consideration participants who endured higher performance losses than the agreed-upon amount but chose to remain in the study for the full duration, since our goal with each participant is to find whether their WTA is higher or lower than the given offer and such participants would clearly accept the same offer for a weaker (i.e. more accurate) slowdown.

\subsection{Does Throttling CPU Frequency Achieve Accurate Slowdowns?}
\label{ssec:throttling}

The experimental protocol hinges on our ability to reliably throttle participants' personal devices by some targeted percentage.
As previously mentioned, we simulate throttled performance by lowering participants' devices' CPU frequency.
While a lowered CPU clock frequency will certainly reduce performance, it was not known a priori if the relationship is linear due to the multi-dimensional factors influencing performance (such as memory capacity, number of cores, cache sizes, disk speed, etc.). 
Given this, is a lowered CPU frequency an appropriate method of simulating throttled device performance?

% Both experimental methodologies rely on throttling CPU frequency by some percentage to achieve the desired slowdown.
% It is important to the experiments' integrity that lowering CPU frequency has the intended effect on lowering overall system performance.
% For example, a potential concern might be that an $X\%$ decrease in CPU frequency might not actually yield an $X\%$ loss in overall performance, which is plausible given that frequency is only one of many different factors contributing to system performance. 
% Given this observation, is a lowered CPU frequency a reasonably precise method of lowering overall system performance, and is it an appropriate means of simulating a performance-throttling patch?

To answer these questions, we benchmarked two Windows 10 devices (a laptop and a desktop device) across three benchmark suites at varying levels of clock frequency:
SPEC2017 intspeed, SPEC2017 intrate, and WebXPRT 3.
The laptop was a Dell XPS 9370 laptop with an Intel Core i7-8550U CPU and 16.0 GB of RAM for the benchmarking, while the desktop device was a custom build PC with an Intel Core i7-8700 CPU with 16.0 GB of RAM and with liquid cooling.
Results are show in Figure~\ref{fig:benchmarks}.
All three benchmarks suites show a highly linear relationship between CPU frequency and benchmark score.
Based on these trends, we find it reasonable to assume that decreasing participants' CPU frequencies by $X\%$ will slow down their devices by roughly $X\%$, and is thus an appropriate method of simulating performance losses.

% Chrome 96
% Balanced: 163
% 100%: 194
% 90% 102
% 80% 91
% 70% 82
% 60% 71

\subsection{Experiment \#2: Simulation Study}

While the incentive compatible study above can product a ``ground truth'' WTA for various performance losses, it is difficult to find participants who are willing to let researchers slow down their personal devices. 
In fact, it took us many months of searching and screening to collect enough participants.
In addition, running this experiment with a large number of participants and for longer periods of time can quickly become costly.
For these reasons, we believe that researchers in this field need alternative methods of estimating users' willingness to accept performance losses.

We now present a simulation-based experiment where participants are briefly subjected to performance throttling on their own personal devices and are then asked to answer questions on their experience. 
The protocol forgoes on incentive compatibility, but still gives participants the chance to actually experience and interact with a throttled device before asking questions.
By forgoing incentive compatibility, we are able to shorten participation from a week or more to about 30 minutes, and also remove the need to subject participants to throttled performance during real-life usage; both of which greatly increase the number of willing participants.

\subsubsection{Experimental Protocol}

Participants begin participation by downloading and running a computer program provided by the researchers.
The program deceives participants into believing that the program is testing unspecified ``features'' on their devices by making some temporary system modifications.
This deception serves two purposes:
First, it gives us informed consent from participants to make temporary modifications to their devices (which is necessary to simulate a performance-throttling patch).
And second, it allows us to change the speed of the participants' devices without their knowledge. 
This is done to avoiding priming participants to think about or notice potential slowdowns in their device during the simulation phase of the experiment.
% The ethics of this deception are discussed in Section \ref{ssec:ethics}.
At this point, the program also tests is ability to actually throttle CPU frequency. 
Participants whose devices cannot be slowed down by the desired amount are removed from further participation. 

After consent is obtained and the device is determined to be eligible for participation, the program asks participants to complete three sets of highly similar tasks (henceforth referred to as ``Task~1'', ``Task~2'', and ``Task~3''), comprised of the following nearly-identical subtasks (only the exact queries requested in steps 2, 3, and 4 above change between tasks):
\begin{enumerate}
    \item Open up Microsoft Word and create a new Word document
    \item Open up a web browser and find the distance in miles between two cities, and add this number to the Word document
    \item Use the a web browser to find an image of a well known landmark, and add this image to the word document
    \item Use the a web browser to find a video of a live music performance on YouTube, and add the URL to the Word document
    \item Export the Word document to a PDF and upload it to a webpage
    \item Close Microsoft Word and the web browser
\end{enumerate}

Participants complete these three tasks back-to-back. 
In between each task, a status bar fills indicating that some unspecified ``features'' are being applied.
Unbeknownst to the participants, the program silently throttles performance during either Task~2 or Task~3 (determined randomly) by setting a cap on CPU frequency during either Task~2 or Task~3.
% \footnote{
%     For example, consider a processor that can run at 2.0 GHz.
%     If the slowdown is set to 30\%, the CPU will be throttled to run no faster than $2.0 \times (1  - 0.3) = 1.4$ GHz.
%     The previously mentioned eligibility screening guarantees that this slowdown is within $\pm 5\%$ of the targeted slowdown.
% }
Device performance is unthrottled during the other two tasks.
After the Task~3 is completed, the program restores the participant's device's performance to its pre-experiment state.
The program then guides participants through an exit survey.

\subsubsection{Exit Survey}

Before eliciting participants' willingness to accept the slowdown they just endured, the exit survey attempts to answer a secondary goal of this work: How noticeable are slowdowns to users?
This is done asking participants try to identify which of Tasks 2 and 3 were throttled.
To incentivize participants to put forth a reasonable amount of effort in answering this question, we offer more money (\$0.25) to participants who can correctly identify which of Tasks 2 or 3 was throttled.

This survey question justifies a significant portion of the experimental protocol:
\begin{enumerate}
    \item It explains the design decision to randomly throttle either Task~2 or Task~3. Our concern was that otherwise the ordering of tasks (i.e. fast-and-then-slow vs. slow-and-then-fast) might have an effect on perceived differences in speed.
    \item It justifies the high degree of similarity between tasks. 
    Making the tasks only minor variations of one another gives participants a fair shot at retroactively comparing device performance between tasks. 
    % For example, if one task had participants stream video content while another had participants run web searches, ranking the tasks would be and apples-to-oranges comparison.
    We also designed the tasks be slight variations of each other (rather than exact replicas) to prevent participants from going into mental ``autopilot'', which might affect perceptions of device performance.
    \item It justifies three sets of tasks instead of the perhaps more obvious choice of only two (i.e. a single fast task and a single slow task). 
    Because the tasks are so similar, it is reasonable to assume the participant will learn the pattern of the tasks after Task~1 and will complete similar subsequent tasks in less time.
    It is possible that a participant might conflate (perhaps even subconsciously) the time it takes to complete a task with their device's performance, which may bias results.
    To avoid this possibility, we use Task~1 as a ``warm-up'' phase to allow participants to become familiar with the tasks and thus increase the likelihood that Task~2 and Task~3 will be completed in roughly the same amount of time.
    Finally, in addition to warming up the participant themselves, we also use Task~1 as a way to warm up the system itself, e.g. by loading system caches.
    
\end{enumerate}

Next, the exit survey reveals to the participant which of Tasks 2 and 3 were throttled and by how much. 
After this disclosure, participants have been fully informed of the true nature of the experiment, and there is henceforth no more deception.

The next section of the survey is a series of yes/no questions designed to elicit the lowest amount a participant would have to be paid to be willing to permanently accept the slowdown they just experienced on their personal device. % TODO need to state that we screen participants to 
We find this minimum dollar amount via a simple variation of the standard exponential search algorithm:

The program first asks participants if they would accept an offer to slow down their device by some percentage in exchange for \$0.
If they accept, there is no lesser minimum and the willingness to accept (WTA) is returned as \$0;
If they decline then the offer price is raised to \$1 and the question is asked again.
If the participant declines again, the offer price is doubled to \$2, then to \$4, and so on for each time the participant declines the offer.
At some point, the offer price will be worth the simulated slowdown\footnote{If the participant \textit{never} accepts the money, their WTA is capped at $2^{14} = \$16384$. We chose this cutoff point since at this point, the payout is multiple times higher than a new high-end system.}.
We now have a price $p$ at which the participant would accept the money over the performance and a price $p/2$ at which the participant would \textit{not} accept the money over the performance.
The participant's minimum WTA therefore lies somewhere in between.
The offer price is then lowered to $\frac{p + p/2}{2}$, or halfway between a known accepted offer price and a known rejected offer price.
Standard binary search then commences, with each offer acceptance setting a new upper bound on the WTA and each offer rejection setting a new lower bound on the WTA. 
Binary search eventually converges on the participant's minimum WTA.
Binary search stops when the difference between the upper and lower bounds converges to \$2, and the WTA is returned as $\frac{(\text{upper bound}) - (\text{lower bound})}{2}$. 

% Once the WTA price point has been found, we ask a few more survey questions.
% These survey questions (and their results) are covered in Section \ref{sec:results}.
% This concludes the survey portion of the experiment.
The program ends by uploading the exit survey responses to a server managed by the researchers.

\subsubsection{Ensuring Data Validity}

To ensure our results our valid, it is necessary to confirm that participants' devices are actually slowed down as intended during the throttled task.
To achieve this, we have the program sample the participants' CPU clock speeds before, during, and after each of the three tasks, and write these clock frequencies to a log file.
Analysis of these log files ensures that participants' devices were actually subjected to the appropriate level of throttling during either Task~2 or Task~3.

\subsection{Experiment \#3: Large-Scale Survey}

While it is easier to recruit participants for simulation study than the incentive compatible study, it is still a challenge to find participants who trust the researchers enough to download a program from the internet and run it with Administrator privileges.
Additionally, it was unexpectedly difficult to find participants whose devices we could reliably slow down by a targeted percentage. 
And last but not least, due to all the work necessary to ensure valid results, neither protocols scale well with larger numbers of participants.
To address these issues, our third methodology is a simple large-scale survey of the questions asked during the simulation-based and incentive compatible studies, but without asking participants to download custom programs or allow background processes to run for days at a time with Administrator privileges. 
In addition to affording larger sample sizes, a survey can help answer the question of whether the incentive compatibility of the first experiment or the performance loss simulation of the second experiment are actually necessary.

\subsection{Ethics and IRB Approval}
\label{ssec:ethics}

All three experimental protocols presented in this work received approval from our IRB.
The main concern was the level of deception in the simulation study, since we slowed down participants' devices without the participants' informed consent.
Due to requirement that we couldn't ``prime'' participants to think about performance or slowdowns, this was a necessary step to achieve valid results.
As required by our IRB, we follow the period of deception with a full debrief explaining that their device was slowed down and why this information was kept hidden from the participants.
Since some participants may object to this deception, we gave participants the opportunity to silently withdraw themselves from the experiment.
As part of the program, we did not receive or consider any data or survey results from participants who silently withdrew from the experiment.

\section{Results}
\label{sec:results}

We now present the results from all three experiments.

\subsection{Experiment \#1: Incentive Compatible Study} \label{sec:results:ssec:exp2}

\begin{figure*}[htb]
    \includegraphics[width=\linewidth]{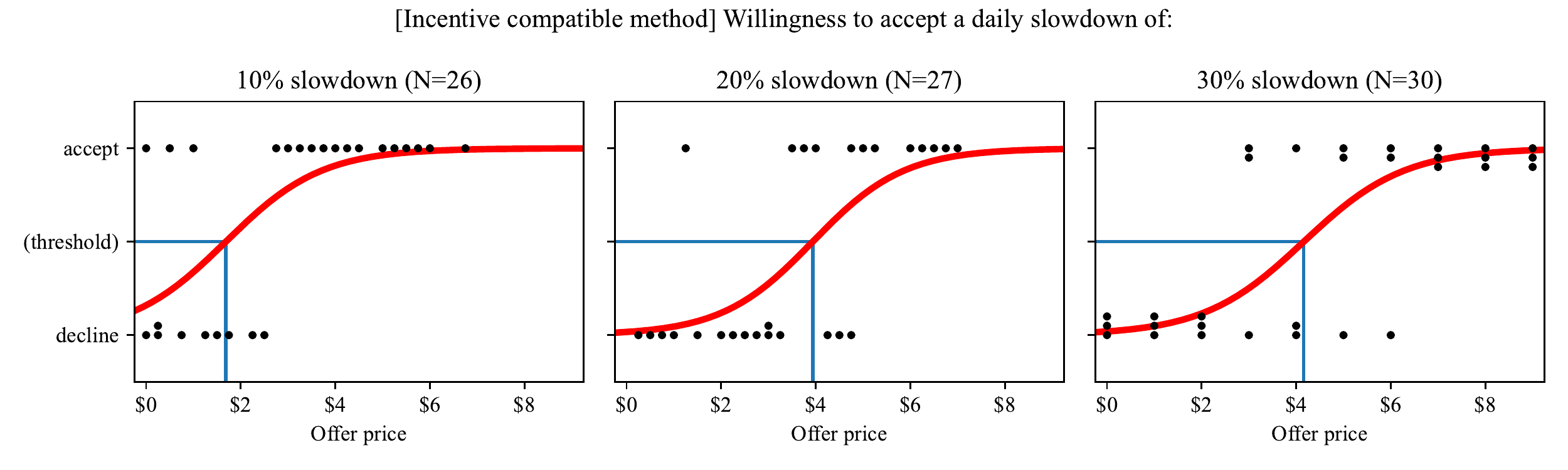}
    \caption{
        Results for the single discrete binary choice question. 
        Each dot represents a unique participant.
        Logistic regression on the data finds that the threshold point (at which point it is more likely than unlikely that a participant will accept a given offer) for slowdowns of 10\%, 20\%, and 30\% is \$2.27 per day, \$4.07 per day, and \$4.44 per day, respectively.
    }
    \label{fig:logreg_study} 

\end{figure*}

\begin{figure*}[htb]
    \centering
    \includegraphics[width=\linewidth]{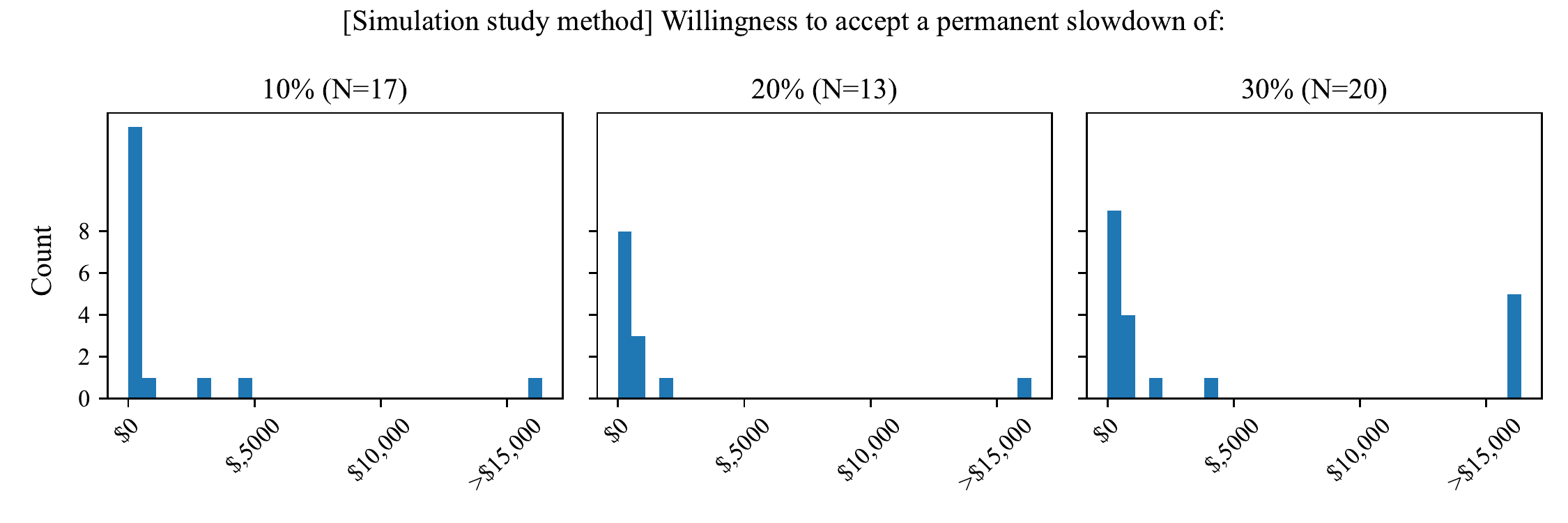}
    \caption{
        A histograms of WTA responses from the simulation study using the exponential search elicitation mechanism.         
        Median WTA values for slowdowns of 10\%, 20\%, and 30\% were \$127, \$169, and \$823, respectively.
    }
    \label{fig:WTASimulationHists}
    \end{figure*}

We ran the incentive compatible study for 7 days at slowdowns of 10\% (N=26), 20\% (N=27), and for 14 days at a slowdown of 30\% (N=30).
For the 10\%, 20\%, and 30\% studies, 10, 16, and 14 participants declined their given offer, respectively, while 17, 13, and 16 participants accepted their offer for the full duration of the experiment, respectively.

The 10\% and 20\% experiments were ran with offer prices ranging from \$0 to \$7 at increments of \$0.25, while the 30\% experiment was ran with offer prices ranging from \$0 to \$9 at \$1.00 increments.

To ensure that participants' computers were actually slowed down for the full duration of their participation, we analyzed the daily log files generated by the program which collected daily samples of device frequency.
We verified that all participants who elected to slow down their computers in exchange for money were actually subjected to the intended slowdown percentage.

For each variant of this experiment, participants are split into one of two groups: 
1) Those who accept the daily monetary offer for the duration of their participation (7 or 14 days), and 2) those who do not.
The first group values their device's performance less than the offer price given to them, indicating that their willingness to accept a 10\%, 20\$, or 30\% slowdown is \textit{less than or equal to} the daily amount offered.

As shown in Figure~\ref{fig:logreg_study} The lowest amount accepted for slowdowns of 10\%, 20\%, and 30\%, was \$0.50, \$3.50, and \$3.00 per day, respectively.

The second group is those who do not accept the daily offer in exchange for throttled system performance.
This group includes those who decline the offer the first time it is given to them as well as those who initially accept the offer but later renege and opt to forgo additional earnings to restore their device's performance.
This second group values their device's performance more than the offer price given to them, indicating that their willingness to accept a 10\%, 20\$, or 30\% slowdown is \textit{greater than} the daily amount offered.
The highest amount declined for slowdowns of 10\%, 20\%, and 30\% was \$2.50, \$4.75, and \$6.00, respectively.

\subsubsection{Finding per-day WTA}

The longitudinal Experiment \#1 yields a two-dimensional dataset with the independent variable $X$ being the offer price and the dependent variable $Y$ being the binary choice of whether not the participant accepted the offer for the duration of the experiment (``Accepts'' are modeled as a $1$ and ``Declines'' are modeled as a 0).
As expected, the results in Figure \ref{fig:logreg_study} show that as the offer price increases, so does the likelihood that a participant will accept their given offer price for a given slowdown percentage. 
We use logistic regression to find the threshold point at which  it is more likely than not that a participant will accept their given offer rather than reject it.

In logistic regression, the log-odds (or logit) is modeled with a linear function:
$$ \text{log}\frac{p(x)}{1 - p(x)} =  \alpha + \beta \cdot x$$.

Since logistic regression has no closed-form solution, we use the STAN modeling framework to build the probability distributions of $\alpha$ and $\beta$ using Bayesian inference. 
We use Stan with 4 Markov chain Monte Carlo (MCMC) chains and $10000$ samples, with a warmup period of $1000$ samples.
Building three models for each of the variants yields the following:
\begin{itemize}
    \item For a 10\% slowdown, our model tells us that  $\alpha$ has a mean of $-5.047$ with a standard deviation of $2.128$ while $\beta$ has a mean of $2.215$ with a standard deviation of $0.844$.
    \item For a 20\% slowdown, our model tells us that  $\alpha$ has a mean of $-9.070$ with a standard deviation of $3.525$ while $\beta$ has a mean of $2.225$ with a standard deviation of $0.835$.
    \item For a 30\% slowdown, our model tells us that  $\alpha$ has a mean of $-5.938$ with a standard deviation of $2.251$ while $\beta$ has a mean of $1.333$ with a standard deviation of $0.473$.
\end{itemize}

From the same Stan model we also find the values of $x$ for which $p(y=1|x) > 0.5$, which we call $x_{\text{crit}}$.
This is the offer price at which it is more likely than unlikely that a participant will accept the offer for the full duration of the experiment.
\begin{itemize}
    \item For a 10\% slowdown, we find $x_{\text{crit}}$ to have a mean of 2.270 with a standard deviation of 0.393. At the 95\% confidence level, the value of $x_{\text{crit}}$ is between 1.536 and 3.016.
    \item For a 20\% slowdown, we find $x_{\text{crit}}$ to have a mean of 4.069 with a standard deviation of 0.355. At the 95\% confidence level, the value of $x_{\text{crit}}$ is between 3.394 and 4.738.
    \item For a 30\% slowdown, we find $x_{\text{crit}}$ to have a mean of 4.443 with a standard deviation of 0.556. At the 95\% confidence level, the value of $x_{\text{crit}}$ is between 3.396 and 5.504.
\end{itemize}

Therefore, it takes an offer of at least \$2.27, \$4.07, and \$4.43 per day for the average user to be willing to accept a 10\%, 20\%, and 30\% drop in performance, respectively.

\subsection{Experiment \#2: Simulation Study}

The simulation study was repeated 3 times, at slowdowns of 10\% (N=17), 20\% (N=12), and 30\% (N=20).

\subsubsection{Finding Device Lifetime WTA}

We plot histograms of participants' willingness to accept slowdowns of 10\%, 20\%, and 30\% in Figure~\ref{fig:WTASimulationHists}.
We cap the maximum reportable WTA at $2^{14} = 16384$, since anything beyond this price far exceeds the cost of a brand-new, very high-end system.

We observe that the majority of participants' WTA price points are grouped towards the lower end with a minority of participants requiring large payouts for any drops in performance.
As expected, greater slowdowns require higher payouts for the slowdown to be ``worth it'' for our participants, although we note that the median WTA for a 20\% slowdown is not much higher than the median WTA for a 10\% slowdown. 
This is likely due to our small sample size.

\subsubsection{How Noticeable are the Slowdowns?}
\begin{figure}[htb]
\center
\includegraphics[width=\linewidth]{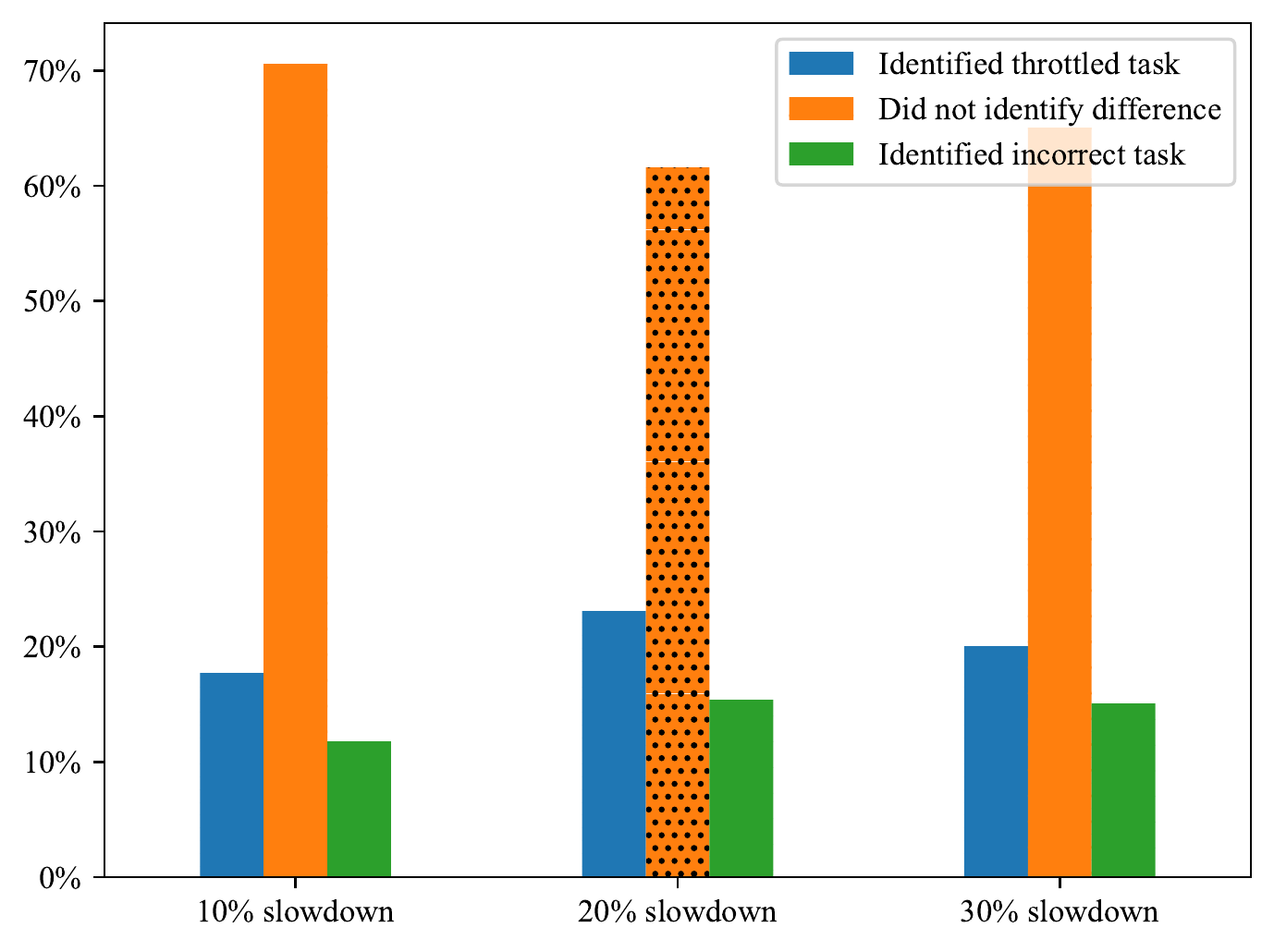}
\caption{Participants in simulation study are split into three groups: Those who correctly identified which of Tasks~2~or~3 was throttled, those who did not report an observed difference in computer speed between Task~2~and~3, and those who incorrectly chose the unthrottled task as the throttled one.}
\label{fig:detection}
\end{figure}

A subgoal of this work was to determine if the stealthily-induced performance throttling in the simulation study was noticeable to participants.
We define a slowdown as being noticeable to a given participant if they were able to correctly identify which of Tasks 2 and 3 were throttled (see Section \ref{sec:methods} for the exact question asked and responses allowed).
Results from this survey question are shown in Figure \ref{fig:detection}.
As expected, participants who endure greater slowdowns are better able to identify which of the two possible tasks was throttled. 
However, participants were, in general, not as perceptive to slowdowns in performance as we expected:
Only about 20\% of participants were able to identify during which task their device was running slower. 
This is likely because the tasks that participants were asked to do are generally not excessively demanding on systems, meaning that the slowdowns may not have had a dramatic effect on system usability.
This might not have been the case if we had asked participants to run compute-heavy tasks instead of simple browser- and word processor-based ones.
Additionally, given that participants were not primed to think about system performance, it is not surprising that they exhibited a generally low level of awareness in this regard;
We expect that participants would be much more perceptive to performance differences if they had been instructed to look for them.

% \FloatBarrier

\begin{figure*} [htb]
    \includegraphics[width=\linewidth]{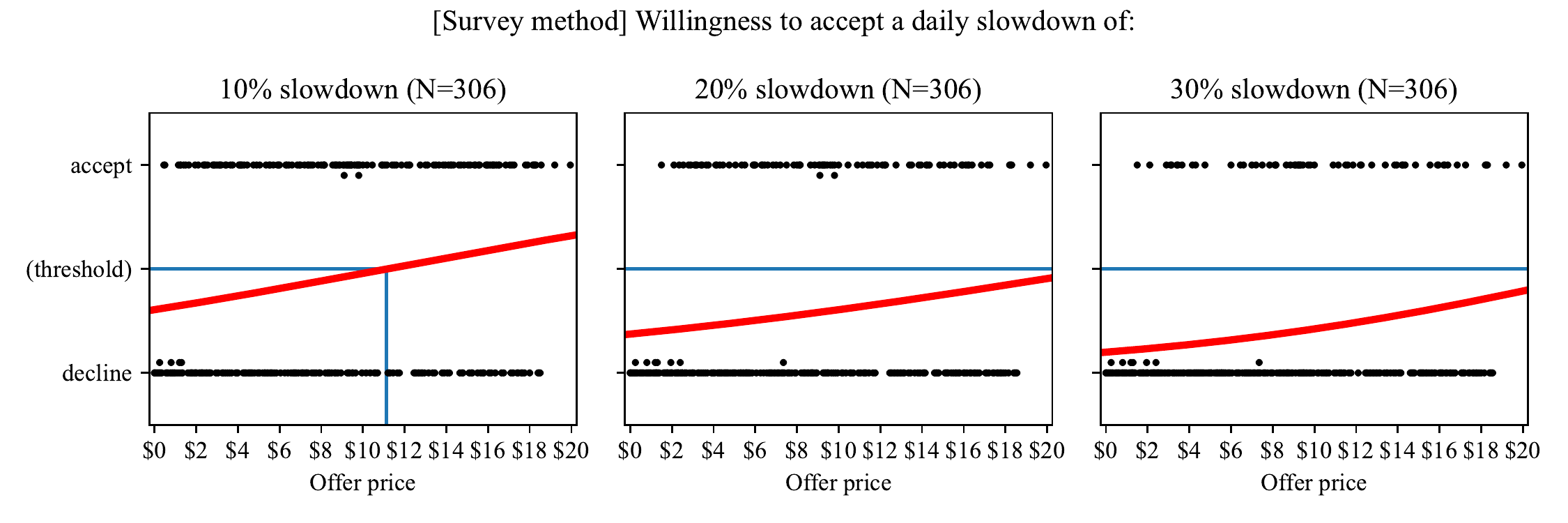}
    \caption{
        A plot of WTA responses from the survey using the single discrete binary choice mechanism.
        Each dot represents a unique participant.
        Compared to the results from the incentive compatible study, the data is much noisier and does not exhibit as clear of trends:
        Logistic regression finds the threshold for accepting a slowdowns of 10\%, 20\%, and 30\% to be \$11.32 per day, \$25.26 per day, and \$27.32 per day.
        This is significantly higher than what was found using the incentive compatible study (\$2.27 per day, \$4.07 per day, and \$4.44 per day, respectively).
    }
    \label{fig:logreg_survey} 
\end{figure*}

\begin{figure*}[htb]
    \centering
    \includegraphics[width=\linewidth]{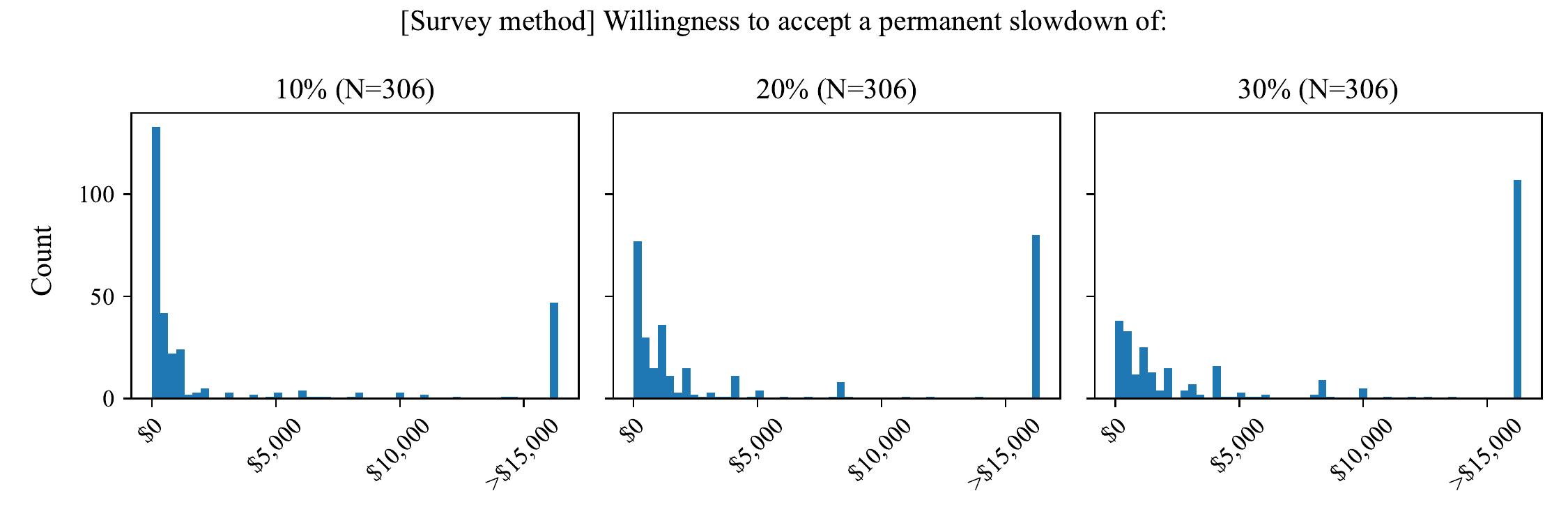}
    \caption{
        A histograms of WTA responses from the survey using the exponential search elicitation mechanism.         
        Median WTA values for slowdowns of 10\%, 20\%, and 30\% were \$499, \$1,214, and \$3,723, respectively.
    }
    \label{fig:WTASurveyHists}
\end{figure*}

\subsection{Experiment \#3: Large-Scale Survey}

We now present results from the large scale survey (N=306). 

\subsubsection{Finding Per-Day WTA}

As part of the survey, each participant was asked the single discrete binary choice question ("Would you accept a permanent $X\%$ loss to your computer's performance in exchange for $\$Y$) with a fixed value of $Y$ but for values of $X=10\%$, $X=20\%$, and $X=30\%$.
Values of $Y$ given to participants ranged between \$0.00 and \$20.00  at \$0.05 increments (some offer prices of $Y$ were given more than once).
Results are shown in Figure~\ref{fig:logreg_survey}.

As with the incentive compatible study, we use the STAN modeling framework to conduct logistic regression on the data.
We use 4 Markov chain Monte Carlo (MCMC) chains and $10000$ samples, with a warmup period of $1000$ samples.
Building three models for each of the variants yields the following:
\begin{itemize}
    \item For a 10\% slowdown, our model tells us that  $\alpha$ has a mean of $-0.836$ with a standard deviation of $0.223$ while $\beta$ has a mean of $0.075$ with a standard deviation of $0.022$.
    \item For a 20\% slowdown, our model tells us that  $\alpha$ has a mean of $-1.485$ with a standard deviation of $0.252$ while $\beta$ has a mean of $0.065$ with a standard deviation of $0.024$.
    \item For a 30\% slowdown, our model tells us that  $\alpha$ has a mean of $-2.217$ with a standard deviation of $0.309$ while $\beta$ has a mean of $0.089$ with a standard deviation of $0.027$.
\end{itemize}

From the same Stan model we also find the values of $x$ for which $p(y=1|x) > 0.5$, which we call $x_{\text{crit}}$.
This is the offer price at which it is more likely than unlikely that a participant will accept the offer for the full duration of the experiment.

\begin{itemize}
    \item For a 10\% slowdown, we find $x_{\text{crit}}$ to have a mean of \$11.31 with a standard deviation of $27.149$. At the 95\% confidence level, the value of $x_{\text{crit}}$ is between $\$7.66$ and $\$15.38$.
    \item For a 20\% slowdown, we find $x_{\text{crit}}$ to have a mean of \$25.26 with a standard deviation of 334.751. At the 95\% confidence level, the value of $x_{\text{crit}}$ is between \$14.69 and \$41.40.
    \item For a 30\% slowdown, we find $x_{\text{crit}}$ to have a mean of \$27.32 with a standard deviation of 54.824. At the 95\% confidence level, the value of $x_{\text{crit}}$ is between \$17.31 and \$39.70.
\end{itemize}

Therefore, according to the survey results, it takes an offer of at least \$11.31, \$25.26, and \$27.32 per day for the average user to be willing to accept a 10\%, 20\%, and 30\% drop in performance, respectively.

\subsubsection{Finding Device Lifetime WTA}

Participants were also asked to give their willingness to accept a device lifetime performance loss using the same exponential search question mechanism used in the Simulation Study. 
We asked participants were asked to give their stated WTA for slowdowns of 10\%, 20\%, and 30\%. 
To improve the data integrity, we removed from the dataset any responses that indicate a WTA of \$0.00 (since this indicates thoughtless or careless responses to the survey questions---it is illogical to accept lowered device performance for nothing in return). 
For the same reason, we also removed from the dataset responses to 10\%, 20\% and 30\% WTA questions that were non-monotonic (e.g. participants who may state they are willing to accept 10\% slowdown for \$200 but also a 20\% slowdown for \$100).
Results are shown in Figure~\ref{fig:WTASurveyHists}.

\section{Discussion} \label{sec:discussion}

A few trends and results from the previous section merit further discussion.

\subsection{Are Surveys a Useful Instrument?}

A major research question of this work is answering the questions, \textit{Are surveys a useful means to estimate users' valuations of performance?}
Our findings indicate that survey are, in fact, \textit{not} a suitable technique for measuring users in this domain.
First, we see that without the presence of incentive compatibility, the daily WTA for slowdowns is significantly higher.
Using the incentive compatible study as a ``ground truth'' measurement of users,  we see that results obtained via the survey are very wide of the mark:
For example, the non-incentive compatible study found the WTA for a 10\% slowdown to be \$11.32 per day---almost $5\times$ higher than the \$2.27 per day as found via the incentive compatible study. 
We find similar error rates  for the 20\% and 30\% slowdown survey responses as well. 
This finding strongly suggests that the presence of incentive compatibility has a significant bearing on results obtained, casting doubt on the validity of results obtained via non-incentive compatible surveys.

We find that the survey also produces much higher results for device lifetime WTA compare to the simulation study. 
For example, the survey found participants' willingness to accept a permanent 30\% loss to their device's performance to be a median value of \$3,723 while the simulation study finds this same value to be only \$823. 
Neither method is incentive compatible, so what might cause this gap?
We see the results in Figure~\ref{fig:detection} to offer an explanation.
A surprisingly low number of participants were able to detect which of the tasks were throttled, suggesting that the throttling did not have a significant impact on user experience, at least during the tasks that were performed.
By experiencing throttled device performance, participants in the simulation study may be more willing to accept permanent device slowdowns (and hence a lower WTA) because they have observed firsthand that throttled device performance is not even necessarily noticeable.

Overall, the conclusion is, unfortunately, that surveys are not a sufficiently accurate instrument in measuring users' valuations of performance. 
Experience-based studies (like our simulation study) offer a step in the right direction, and ultimately incentive compatible study designs may be a necessity.

\subsection{Cross-Validating the Experiments} 

Our experiments aim to capture users' willingness to accept performance slowdowns, but expressed two different ways:
The incentive compatible study measures \textit{per day} WTA, whereas the simulation study measures \textit{per device lifetime} WTA.
(The survey measures both per day WTA and per device lifetime WTA).
Although different, both measures are useful in their own way:
Per day WTA puts a price to performance losses that are temporary or reversible (e.g. optional security features like disabling hyperthreading to prevent speculation attacks).
In contrast, per device lifetime WTA captures the user's value of performances losses that are permanent and irreversible (e.g. patchesto Meltdown  or the so-called ``Batterygate'' issue with Apple iPhones).
Both scenarios are common enough that they each warrant independent investigation.

% Section on trying to cross validate them since only the per day is incentive compatible
That said, finding the per device lifetime WTA is perhaps more necessary than finding per day WTA because users' value of performance is more pertinent when performance losses are \textit{not} optional, which, historically speaking, have been permanent and irreversible.
However, the per device lifetime WTA we elicit using the simulation study is not incentive compatible, nor can it easily be converted into an incentive compatible experimental design
\footnote{The only option we are aware of here is to pay study participants to permanently throttle their device's performance, which would be both expensive and difficult to enforce.}.
Since the lack of incentive compatibility can skew or adulterate experimental results, it would be useful if we can use the results from the incentive compatible study to validate the results of the simulation study.
A reasonable approach might be to divide the device lifetime WTA by the per day WTA and check if the resulting number (in terms of days) approximates a reasonable device lifetime:
We find that for slowdowns of 10\%, 20\%, and 30\%, this simple cross-validation calculation yields results of 56, 43, and 186 days, which are shorter than most device lifespans.
This may be explained by the lack of incentive compatibility in the simulation study, or perhaps because users are not as rational when making decisions about long term events (like permanent performance loss).

\section{Limitations}
\label{sec:limitations}

We now discuss some of the limitations of our methodology and results.

{\bf Sample Demographics:} A potential concern is that our sample
population---drawn from Mechanical Turk workers---is not representative
of the population at large.  We ran the simulation study with a 30\%
slowdown on a second sample population---collected from our research
institution---and find that the results generally corroborate each
other.  We find that the second sample population's willingness to
accept a 30\% slowdown in performance is \$1004
(N=17), higher than the Mechanical Turk median of \$823.
We suspect that the results from Mechanical Turk are much more generalizable than 
the results collected from our own research institution based on conclusions from prior work \cite{redmiles2019well}.

{\bf Fidelity:} One caveat for both experiments is that we were not
able to throttle device performance with surgical precision.
For example, in the simulation study we
had to allow for a $\pm 5\% $ tolerance band in order to find enough
participants with eligible devices.
In the incentive compatible study, the throttled frequency for a 
few participants in the 30\% as recorded in the daily log files was
slightly higher than the desired 30\% slowdown target for unknown reasons.

{\bf Devices:} A final limitation is that these experiments were
conducted on Windows 10 desktops and laptops only.  We limited our
studies to Windows 10 since it was the only major operating system
on which we could reliably throttle performance. While it
is possible to control CPU clock frequency on Linux devices, we
opted to recruit from Windows users instead since the larger user
base makes it easier to recruit willing participants. In contrast,
we could not successfully throttle MacOS devices, and found that
throttling phone performance would require device-level root access,
which we suspect many users would find to be unacceptable.

\section{Applications} \label{sec:applications}

In addition to defining the ``exchange rate'' between performance and user satisfaction,
this work has broad implications for hardware security.  Many
day-to-day computer buyers are largely unfamiliar with vulnerabilities
like Rowhammer and Spectre.  Due to this lack of knowledge---aka
information asymmetry---they are not able to appropriately place
a value on hardware security features.  This produces the so-called
Market for Lemons problem, where better quality (in this case, more
secure) goods are driven away from the marketplace because consumers
are unwilling to pay a premium for features they cannot
identify~\cite{akerlof1978market}.  A consequence is that hardware
companies are unwilling the adopt security features unless the
overheads are razor-thin~\cite{hastings2020wac}. This severely
limits the range and scope of security features that are adopted
in real-world products, and leaves many real threats un- or
under-patched for fear of causing harm to performance~
\cite{kocher2019spectre,kim2014flipping}.

However, our results indicate that perhaps there is more room for
security features than previously thought.  We point to two
observations: First, during the incentive compatible study, all but
one of the participants who accepted their offer participated for
the full duration of the experiment.  This indicates that after
participants accepted their first daily offer, their experience
with their throttled device was, in general, not sufficiently bad
enough to warrant removing themselves from the experiment and
forgoing additional earnings.  Second, the WTA as found via the
simulation study are lower than the WTA found via the survey,
suggesting that the experience of throttled performance is not as
bad as users expect.  Combined, this suggests that participants'
high resistance to performance losses is more psychological than
based on actual needs.  While these experiments need to be conducted
for servers where there is less information asymmetry, the takeaway
for end user devices is that user resistance to performance losses
due to security patches and updates may be artificially holding
back the deployment of security features.

\section{Related Work}
\label{sec:rw}

\textbf{User-Centered Design:}
In the architecture and systems communities, the end user is often seen as being many layers of abstraction removed from the hardware, and thus the end user oftentimes is not considered  during the hardware and systems design process.% in the same way that other metrics like throughput or performance per watt are, perhaps because users' preferences and values are not easily measured.
Our work and other similar lines of research attempt to cut through these layers of abstraction by providing user-centered metrics to better aid user-focused system design. 
Several user studies have leveraged dynamic voltage and frequency scaling (DVFS) to help balance the competing demands of energy efficiency and user satisfaction~\cite{mallik2008piscel,shye2008power,shye2009wild,shye2008learningleveraging}. 
Individualized quality of service (QoS) metrics have also been proposed as a means towards achieving this balance~\cite{yan2015characterizingqoe,yan2016redefiningqos,yan2019improvingenergyefficiency}.
Other work identifies the components and design configurations that yield higher user satisfaction~\cite{yan2019bridgingmobile,halpern2016risetopower}.

User-centered metrics for improving user satisfaction have not been confined to academia: Intel's Project Athena has introduced metrics for its EVO line of laptops called Key Experience Indicators (KEIs) that quantify elements of the user experience, such as wake time, responsiveness, and charging times~\cite{intel2020athena}.

\textbf{Incentive Compatible Mechanisms:}
% Incentive compatibility mechanisms 
While the above work on user-centered design is thematically similar to our own, the methodologies do not employ incentive compatible study designs and are subject to the same common pitfalls as our simulation-based and survey-based methodologies.
Our work aims to raise the bar for user studies for hardware design by introducing incentive compatible methodologies and mechanisms.
In particular, take inspiration from previous incentive compatible studies, and in particular from Brynjolfsson, Collis, and Eggers, who conducted incentive compatible experiments to determine the value created by free online digital goods (such as access to web searches, online maps, social media, etc.)~\cite{brynjolfsson2019using}; this work produces incentive compatible results via mechanisms such as Becker-DeGroot-Marschak~(BDM) lotteries \cite{becker1964measuring}, Best-Worst~Scaling~(BWS)~\cite{louviere2015best},and single discrete binary choice~(SDBC) experiments~\cite{carson2014consequentiality}.
We use the SDBC mechanism for achieving incentive compatibility in our own work but note that other mechanisms may be useful for future experiments on consumers' willingness to pay for system features like performance or security.

\textbf{Pricing Performance:} Measuring the value of performance in terms of dollars has been attempted before, but from the perspective of businesses rather than end users.
Studies performed by web companies have found that increases to latency hurt revenues \cite{glinden, oreilly09}, largely because keeping users engaged with services requires low latency~\cite{nielsen1994usability}.
% Amazon found a 1\% drop in revenue for every 100ms of latency\footnote{It should be noted that this oft-repeated claim is not backed by any published work known to the authors.}\cite{glinden}.
% Bing also reported that a 0.5 second delay per web request reduced revenue per user by 1.2\% while a 2 second delay per web request reduced revenue per user by 4.3\% \cite{oreilly09}.
% In general, it has long been known that keeping users engaged requires low latency~\cite{nielsen1994usability}.
% Outside of the cost of losing users due to slowdowns, Barroso analyzes the price of performance for web companies in terms of hardware cost and the cost of power, but does not measure consumers' valuation of performance \cite{barroso2005price}.
The value of latency has also been studied in financial markets and high-frequency trading~\cite{moallemi2010cost, riordan2012latency}.
% In datacenters, performance has long been measured in terms of performance per watt \cite{barroso2005price}.
To our knowledge, our work is the first of its kind to put a price to end user's value of performance.

\section{Conclusion} \label{sec:conclusion}

This paper presents the first study on hardware behavioral economics.
Three methodologies are used to elicit users' willingness to accept performance losses.
Our first incentive compatible experiment finds users' WTA to accept performance losses of 10\%, 20\%, and 30\% to be \$2.27, \$4.07, and \$4.43 per day, respectively.
The second experiment, while not incentive compatible, still allows participants to experience a loss of performance on their personal device, and finds the WTA to be \$127, \$169, and \$823 for slowdowns of 10\%, 20\%, and 30\%, respectively.
Finally, our survey-based study finds these same numbers to be \$11.32, \$25.26, and \$27.32 per day and $\$499$, $\$1214$, and $\$3723$ overall for performance losses of 10\%, 20\%, and 30\%, respectively.
Importantly,  we find that the survey-based study is not a suitably accurate method of determining users' WTA of performance losses. 

We are in the midst of a revolution in computer architecture and
computer hardware design~\cite{turingaward}.  The huge demand for
vertically integrated products (ranging from iPhones to cars to
push-to-order IoT widgets), along with the rise of open source
hardware, has pushed hardware companies to rapidly innovate to create products that must meet high
demands of integration, performance, energy efficiency,
and cost requirements. 
%  While some of these conditions have been
% present for several years (leading to the creation of significant
% security risks, as evidenced by the recent glut of hardware
% vulnerability reports), we expect these vulnerabilities to become
% more commonplace going forward.
In the push to meet the changing design requirements, we introduce a new metric---willingness to accept performance losses---that 
defines the ``exchange rate'' between user satisfaction and system performance.
This new metric lets systems designers and architects to quantitatively consider the dollar cost, from the users' perspective, of performance-costing features when designing future systems.

%%%%%%% -- PAPER CONTENT ENDS -- %%%%%%%%

%%%%%%%%% -- BIB STYLE AND FILE -- %%%%%%%%
\bibliographystyle{IEEEtranS}
\bibliography{refs}
%%%%%%%%%%%%%%%%%%%%%%%%%%%%%%%%%%%%

\end{document}